# Perovskite Phase Heterojunction Solar Cells

*Ran Ji,[1,2] Zongbao Zhang,[1,2] Yvonne J. Hofstetter,[1,2] Robin Buschbeck,[1,2] Christian Hänisch,[1] Fabian Paulus[2] and Yana Vaynzof[1,2*]*

[1] Integrated Centre for Applied Physics and Photonic Materials, Technische Universität Dresden, Nöthnitzer Str. 61, 01187 Dresden, Germany

[2] Center for Advancing Electronics Dresden (cfaed), Technische Universität Dresden, Helmholtzstraße 18, 01089 Dresden, Germany

**Corresponding Author**

*Yana Vaynzof, e-mail: yana.vaynzof@tu-dresden.de

**Abstract**

Modern photovoltaic devices are often based on a heterojunction structure where two components with different optoelectronic properties are interfaced. The properties of each side of the junction can be tuned by either utilizing different materials (e.g. donor/acceptor) or doping (e.g. PN Si junction) or even varying their dimensionality (e.g. 3D/2D). In this work we demonstrate the concept of phase heterojunction (PHJ) solar cells by utilizing two polymorphs of the same material. We demonstrate the approach by forming $\gamma$-CsPbI$_3$/$\beta$-CsPbI$_3$ perovskite PHJ solar cells. We find that all of the photovoltaic parameters of the PHJ device significantly surpass those of each of the single-phase devices, resulting in a maximum power conversion efficiency of 20.1%. These improvements originate from the efficient passivation of the $\beta$-CsPbI$_3$ by the larger bandgap $\gamma$-CsPbI$_3$, the increase in the built-in potential of the PHJ devices enabled by the energetic alignment between the two phases and the enhanced absorption of light by the PHJ structure. The approach demonstrated here offers new possibilities for the development of photovoltaic devices based on polymorphic materials.

Most photovoltaic technologies rely on the use of a junction to enable their function as an efficient solar cell.[1–5] The fundamental concept behind this approach is independent of how the junction is realized and is based on the combination of two components with different optoelectronic properties. The simplest realization of a heterojunction is the combination of two different materials. For example, organic solar cells rely on the formation of a junction between donor and acceptor organic materials in a 'bulk-heterojunction', thus facilitating charge separation at the heterointerface (Fig. 1a).[2,6,7–9] The well-established Si solar cells are based on a different type of junction, one that is formed by p- and n-doping the Si crystal on either side of the homojunction (Fig. 1b).[10] However, in CdTe[11], CIGS[12], and GaAs[13] solar cells, a PN junction can also be formed as a heterojunction (Fig. 1c). Finally, another noteworthy example is the use of junctions of varying dimensionality, such as a 3D/2D junction (Fig. 1d) in perovskite solar cells,[14] leading to improvements in their efficiency and stability.

Polymorphism is observed in many classes of crystalline materials.[15–19] This phenomenon is based on the ability of a material to exist in several crystalline phases, with different structural arrangement of atoms or molecules.[15,18] Both organic and inorganic materials may exhibit polymorphism which has been shown to strongly impact their optoelectronic properties.[20] It is thus interesting to explore the possibility to combine two polymorphs of the same material to form a phase heterojunction.

A promising candidate for the demonstration of this concept is the material class of metal halide perovskites (MHPs). MHPs exhibit favourable optoelectronic properties, which have led to their successful integration in high performance photovoltaic devices.[21–25] Moreover, MHPs exhibit multiple polymorphs based on their composition. For example, the inorganic CsPbI$_3$ perovskite exhibits four different polymorphs,[26] three of which are photoactive and have been exploited for the fabrication of photovoltaic devices.[27] α-CsPbI$_3$ can be formed by spin-coating a 1:1 CsI:PbI$_2$ solution in N,N-dimethylformamide (DMF) and heating to 335 °C.[28] With cooling to room temperature, the perovskite would not directly turn to yellow phase (δ-CsPbI$_3$). Instead, CsPbI$_3$ tends to form the metastable polymorphs, where the cubic phase (α-CsPbI$_3$) with high-symmetry distorts initially turns to a tetragonal phase (β-CsPbI$_3$) followed by a further change to the orthorhombic (γ-CsPbI$_3$).[30] Notably, although β-CsPbI$_3$ and γ-CsPbI$_3$ are both metastable phases, their optical bandgap are clearly different (β-CsPbI$_3$≈1.68 eV, γ-CsPbI$_3$≈1.74 eV).[29] Additionally, their lattice constants also change as the relative orientation of the [PbI$_6$]$^{4-}$ octahedra is altered.[30] With the development of a stabilizing strategy, all three of the photoactive polymorphs can be maintained at room temperature. α-CsPbI$_3$ or γ-CsPbI$_3$ solar cells were reported to achieve a power-conversion-efficiency (PCE) of ~17%.[31,32] β-CsPbI$_3$ solar cells result in higher PCEs than α- and γ-CsPbI$_3$ based devices, with high short-circuit current densities (J$_{SC}$) of over 20 mA/cm$^2$ owing to their narrower bandgap.[29,33–36] The PCEs of β-CsPbI$_3$ based solar cells have surpassed 20% and 18% in standard and inverted structure, respectively.[36,37]

To realise the PHJ concept, we chose to combine the β- and the γ-phases of CsPbI$_3$. We find that compared with single-phase (SP) solar cell devices, the PHJ devices exhibit a simultaneous increase in all of the photovoltaic parameters, reaching a maximum fill factor of 84.17% and power conversion efficiency of 20.17%. Microscopic and spectroscopic characterisation revealed that this increase in performance is a result of effective defect passivation, beneficial energetic alignment and enhanced light absorption in the PHJ structures.

**Fabrication of the PHJ and structural characterisation**

Our choice of the β-CsPbI$_3$ and γ-CsPbI$_3$ phases for the demonstration of the PHJ concept was motivated by their relatively low temperature processing, unlike the *α*-CsPbI$_3$ that requires annealing at >300 ºC.[28] Considering that solution processing of all polymorphs of CsPbI$_3$ requires the use of the same type of polar solvent,[38] we utilised a hybrid deposition approach that combines both solution processing and thermal evaporation. Specifically, we first deposited the narrower bandgap β-CsPbI$_3$ by a single step solution deposition method,[29] followed by vapour deposition of the γ-CsPbI$_3$ (Fig. 2a). The β-CsPbI$_3$ was fabricated following the 'HPbI$_3$' (dimethylammonium iodide, DMAI) method. DMAI is an effective volatile additive that allows to manipulate the crystallization process of β-CsPbI$_3$ by forming an intermediate phase.[22] The additive is then sublimated during the annealing process. The γ-CsPbI$_3$ was fabricated by co-evaporation process we developed in a previous study.[39] In this process, a small amount of phenethylammonium iodide (PEAI) is co-evaporated alongside the perovskite precursors and serve as the additive and stabilizer to maintain the γ-CsPbI$_3$. Throughout our study we kept the thickness of the solution-processed β-CsPbI$_3$ fixed at 370 nm and varied the thickness of the vapour deposited γ-CsPbI$_3$ from 10 nm to 200 nm.

To characterise the microstructure of the deposited layers, their surface was imaged using scanning electron microscopy (SEM). The solution-processed β-CsPbI$_3$ shows a large and uniform grain size with an average diameter of ~300 nm. The microstructures of PHJ10 and PHJ20 are similar, however with some additional small domains visible at the grain boundaries and on the surface of the β-CsPbI$_3$ grains. Increasing the thickness of γ-CsPbI$_3$ further leads to the formation of grains with comparable sizes to those of the underlying β-CsPbI$_3$, however with orderly striped edges on the surfaces of the grain, particularly evident for the PHJ100 films. This is likely to be associated with the highly oriented growth of γ-CsPbI$_3$ along the (0 0 *l*) planes, as shown in our previous work and Supplementary Figure 1a.[39] With the thickness of the top γ-CsPbI$_3$ increasing to 200 nm, the surface morphology becomes more disordered with a myriad of grains with a range of shapes and sizes. A comparison to a 200 nm thick γ-CsPbI$_3$ film deposited on the glass reveals that such a film exhibits far smaller grains than the β-CsPbI$_3$ or the PHJ films with thicknesses <200 nm. This suggests that the large grains of the underlying β-CsPbI$_3$ impact on the growth of the γ-CsPbI$_3$ perovskite, such that the grain size of the latter is increased when deposited on a perovskite layer. This is further confirmed by SEM images of γ-CsPbI$_3$ of varying thicknesses deposited on glass substrates (Supplementary Fig. 1b-f). Even the thinnest 10 nm thick γ-CsPbI$_3$ samples exhibit a compact film with small grains, which are hard to distinguish due to the lack of contrast at the boundaries. While the grains become slightly larger with increasing thickness of γ-CsPbI$_3$, they remain significantly smaller than those observed for the PHJ samples, confirming that the growth is influenced by the underlying β-CsPbI$_3$ microstructure.

To investigate the composition of the small clusters observed on top of the large grains of the PHJ samples, we performed energy-dispersive x-ray spectroscopy (EDX) elemental mapping on the PHJ100 sample (Supplementary Fig. 2a-d). The experiment revealed that the distribution of Pb, Cs and I is fully homogeneous across the layer, suggesting that these small clusters are CsPbI$_3$ domains formed during the evaporation process.

Cross-sectional SEM imaging was performed to examine the horizontal microstructure of the PHJ100 film (Fig. 2c). The cross-section shows that the β-CsPbI$_3$ layer exhibits large vertical grains with few boundaries in the direction of transport. Meanwhile, the 100 nm thick γ-CsPbI$_3$ perovskite layer is comprised of similarly sized grains, with a very clear boundary between the two layers. For comparison, the cross-sectional SEM images of single β-CsPbI$_3$ and γ-CsPbI$_3$ layers are shown in Supplementary Fig. 1a, b, respectively. As has been already observed from surface images, the growth of γ-CsPbI3 on top of the β-CsPbI3 leads to far larger grain sizes than when deposited on glass, suggesting that the microstructure of the γ-CsPbI3 is templated by that of the underlying β-CsPbI3 layer.

To determine the crystal structure of the perovskite films, we used x-ray diffraction (XRD) techniques, including grazing incidence x-ray diffraction (GIXRD). We indexed the XRD patterns to β-CsPbI3 and γ-CsPbI3 (Supplementary Fig. 3a).[26,40,41] The obvious difference in the peak positions between the β and γ phases can be observed, and is especially evident for the ~28° peak (Fig. 2d). The vapour deposited γ-CsPbI3 grows along the (0 0 $l$) planes, leading to strong diffraction peaks of the (0 0 2) and (0 0 4) reflections, which appear at lower 2θ angles than the diffraction peaks of the β-CsPbI3, which correspond to the (1 1 0) and (2 2 0) reflections. With increasing thickness of the PHJs, contributions from both phases can be observed, especially noticeable in the PHJ100 and PHJ200 films. The full XRD patterns can be seen in Supplementary Fig. 3b.

To further confirm the structure of the $\gamma$-CsPbI3/$\beta$-CsPbI3 PHJ, GIXRD experiments were performed on the PHJ100 sample by varying the incidence angle Ω from 0.4° to 5° (Fig. 2e). The lower incidence angles are more surface sensitive, and the XRD patterns (measured around ~28°) show a stronger contribution of $\gamma$-CsPbI3 (0 0 4) diffraction peak, with a smaller contribution of the $\beta$-CsPbI3 (2 2 0) peak. Increasing Ω to 5° makes the measurement more bulk sensitive, leading to a far stronger signal from the $\beta$-CsPbI3 (2 2 0) peak. These measurements confirm the formation of a $\gamma$-CsPbI3/$\beta$-CsPbI3 PHJ that retains the phase of each of the junction layers.

**Optical properties of the PHJ**

To investigate the optical properties of the perovskite films, the absorption and photoluminescence (PL) properties were characterised. Fig. 3a displays the UV-vis absorption spectra of the SP and PHJ films. The absorption edges of the γ- and β-CsPbI$_3$ are ~710 nm and ~740 nm, respectively, which is consistent with previous reports in the literature.[29,32,34,36,41] These measurements further confirm that the phases of the CsPbI$_3$ films deposited by solution and evaporation indeed correspond to the β and γ phases. To precisely evaluate the bandgap of each of the two phases, Tauc plots (Supplementary Fig. 4) were extracted from the absorbance data. The bandgap of γ- and β-CsPbI$_3$ were found to be ~1.74 eV and ~1.69 eV, respectively. Their two slightly different bandgaps result in the absorbance spectra of the PHJ samples displaying features originating from both phases, especially evident in the PHJ200 spectrum that exhibits a strong absorption feature around 680 nm.

Steady-state PL measurements are shown in Fig. 3b. The PL peak positions of γ- and β-CsPbI$_3$ are at ~710nm and ~740 nm, respectively, which is consistent with their absorption edges. The PL spectra of the PHJ samples are centred around 740 nm, consistent with light emission only from the low bandgap β-CsPbI$_3$. It is curious that no emission from the γ-CsPbI$_3$ is observed, even for the PHJ samples with the thickest γ-CsPbI$_3$ layers. There are two things to consider when examining the PL behaviour. First, the bandgap of the γ-CsPbI$_3$ is larger than that of the β-CsPbI$_3$ which would enable energy transfer from the γ-CsPbI$_3$ to the β-CsPbI$_3$, thus supressing emission from the γ-CsPbI$_3$. Secondly, the energetic alignment between the two phases may lead to charge transfer. Indeed, based on ultra-violet photoemission spectroscopy (UPS) depth profiling measurements (Supplementary Fig. 5), the valence bands of the β-CsPbI$_3$ and γ-CsPbI$_3$ phases of the junction are aligned at the heterointerface, which suggests that the conduction band of the β-CsPbI$_3$ is slightly lower than that of the γ-CsPbI$_3$. Upon photoexcitation, this would result in electron extraction towards the β-CsPbI₃, while the holes remain in the γ-CsPbI₃, thus serving as an additional mechanism that prevents radiative recombination in the γ-CsPbI₃.

Monitoring the intensity of the PL and the corresponding trend in PL quantum yield (PLQY) (Fig. 3c) reveals an interesting trend. With the introduction of 10 nm of γ-CsPbI₃ on top of the β-CsPbI₃, the PLQY increases by a factor of two, suggesting a thin layer of γ-CsPbI₃ can effectively passivate trap states at the surface of the β-CsPbI₃, thus suppressing non-radiative recombination. However, with the further increase in the thickness of the γ-CsPbI₃ layer, the PLQY gradually decreases up to the value of the SP γ-CsPbI₃. We believe that this trend is a result of electron transfer from the γ-CsPbI₃ to the β-CsPbI₃ phase across the heterointerface, which opens an additional relaxation pathway via an exciton-electron annihilation in the β-CsPbI₃. As the thickness of the $\gamma$-CsPbI₃ increases, a larger number of electrons formed upon a photoexcitation of the $\gamma$-CsPbI₃ can be transferred to the $\beta$-CsPbI₃, thus progressively lowering the PLQE of the $\beta$-CsPbI₃.

Fig. 3d displays the time-resolved photoluminescence (TRPL) decay curves of SP and PHJ films. Considering that the curves exhibit a non-monoexponential decay, the lifetime was evaluated using $\tau_{ave}$, with the details of the fitting process and the resulting parameters summarised in Supplementary Fig. 6 and Supplementary Table 1. We observe that the $\tau_{ave}$ of β-CsPbI₃ is 9.5 ns, which is consistent with other reported values in the literature.[34,37] However, γ-CsPbI₃ exhibits a much longer $\tau_{ave}$ up to 125 ns. With the introduction and increasing thickness of γ-CsPbI₃ in the PHJ, the $\tau_{ave}$ increases from 9.56 ns of β-CsPbI₃ to 104 ns of PHJ100. However, the $\tau_{ave}$ of PHJ200 decreases to 77 ns. These results are consistent with our previous observation that γ-CsPbI₃ can effectively passivate the underlying β-CsPbI₃.

**Photovoltaic performance of SP and PHJ solar cells**

To investigate the photovoltaic performance of the SP and PHJ layers, they were incorporated into photovoltaic devices with an inverted architecture following the structure: glass/ITO/MeO-2PACz/β-CsPbI$_3$/γ-CsPbI$_3$/PCBM/BCP/Ag (Fig. 1e). Additionally, SP devices with the optimal thickness (370 nm for β-CsPbI$_3$ and 500 nm for γ-CsPbI$_3$) were fabricated. The photovoltaic performance parameters of the devices are summarised in Fig. 4a-d. Comparing the performance of the SP solar cells reveals that the open-circuit voltage ($V_{OC}$) of the β-CsPbI$_3$ SP devices is approximately 0.1 V lower than that of the γ-CsPbI$_3$ SP solar cells, a much higher difference than the 0.05 eV change in their bandgaps. This suggests that additional non-radiative losses are present in the β-CsPbI$_3$ solar cells, in agreement with the short PL lifetime of the corresponding thin films and with the previous reports of a high density of surface defects in the case of β-CsPbI$_3$ solar cells.[29] On the other hand, the short-circuit current density ($J_{SC}$) of the β-CsPbI$_3$ devices is substantially higher than the $J_{SC}$ of γ-CsPbI$_3$ devices. This is consistent with the lower bandgap of β-CsPbI$_3$ which enables absorption across a larger spectral region, thus resulting in a higher photocurrent. Both SP devices result in a similar range of fill factors (FF), with slightly higher values in the case of the γ-CsPbI$_3$ solar cells. The opposing trends of the $V_{OC}$ and $J_{SC}$ between the two SP devices lead to a similar power conversion efficiency (PCE) in both cases of up to 15%.

The PHJ devices show a gradual increase in all of the photovoltaic parameters, reaching an optimum at PHJ100 (i.e. 100 nm γ-CsPbI$_3$). PHJ100 devices exhibit a high $V_{OC}$ of up to 1.15 V, a $J_{SC}$ reaching up to 20 mA/cm$^2$ and a high FF of 82%. The simultaneous increase in all the photovoltaic performance parameters leads to a significant increase in the PCE of PHJ devices, surpassing 19% in the case of PHJ100 (for maximum power point tracking see Supplementary Fig. 7). Considering that the PL studies indicated that thin layers of γ-CsPbI$_3$ may passivate surface defects of the underlying β-CsPbI$_3$, we fabricated devices in which ultra-thin layers of γ-CsPbI$_3$ were evaporated (2 nm and 5 nm). Since passivation is a surface effect, such thin layers enable to identify to which degree passivation is responsible for the increase in the photovoltaic performance. We observe that the evaporation of 2 nm of γ-CsPbI$_3$ results in an increase in $V_{OC}$ and FF, thus increasing the average PCE of the β-CsPbI$_3$ SP devices from 13.2% to 14.3%. This moderate increase is consistent with surface passivation and is similar to what is observed when other surface passivation methods are applied (Supplementary Fig. 8). This result and the fact that the $V_{OC}$ and FF continue to increase up to PHJ100 suggests that while passivation of the surface defects of the β-CsPbI$_3$ contributes to the increase in photovoltaic performance of the PHJ, it is not the sole factor leading to an enhancement in performance. Examining the energy band diagram reveals that the resultant energetic alignment increases the built-in potential of the device. Considering that all of the devices are fabricated with the same extraction layers, and that these extraction layers are always pinned a certain energetic distance below/above the conduction/valence band (Supplementary Fig. 9, left), the extraction of electrons from the higher lying conduction band of the γ-CsPbI$_3$ results in an increase in the built-in potential ($V_{BI}$), and consequently the $V_{OC}$ and FF. Such a strategy has been previously proposed by device simulations, in which higher bandgap layers were introduced near the extraction layers, thus resulting in an increase in the $V_{BI}$.[45] Importantly, this highlights that with the structure β-CsPbI$_3$/γ-CsPbI$_3$ the observed increase in $V_{OC}$ and FF would only occur for inverted architecture devices, since introducing the PHJ in a standard device architecture would lead to the opposite effect, in which the $V_{BI}$ is reduced (Supplementary Fig. 9, right). Indeed, when we fabricated devices with the structure Glass/ITO/TiO$_2$/PVK/Spiro-OMeTAD/Au (PVK = β-CsPbI$_3$ or β-CsPbI$_3$/γ-CsPbI$_3$) the PHJ devices showed a significantly decreased performance, in particularly strong reduction in the $V_{OC}$ and FF (Supplementary Fig. 10). To utilize the PHJ concept in standard architecture devices would thus require to reverse the order of layers, i.e. γ-CsPbI$_3$/β-CsPbI$_3$, however this is not yet possible using existing fabrication protocols, since the deposition of the β-CsPbI$_3$ occurs from solution and would thus dissolve the underlying γ-CsPbI$_3$. In the future, we plan to develop protocols to deposit β-CsPbI$_3$ by thermal evaporation, which would enable to form PHJ at any order and thus introduce them into standard architecture devices as well.

The increase in the J$_{SC}$ observed when thicker layers of γ-CsPbI$_3$ are used is associated with the increased absorption of the PHJ devices as compared to the SP ones. This is evident from comparing the external quantum efficiency (EQE) spectra of the devices (Fig. 4e), which confirm the far higher contribution to photocurrent in the range 600 nm to 740 nm of the PHJ devices as compared to either of the SP solar cells. We highlight that only the γ-CsPbI$_3$ device exhibits a different onset of the EQE spectra, due to its larger bandgap. All of the PHJ devices show the same onset as that of the β-CsPbI$_3$ only devices, with an increase in the 600-740nm range due to the additional absorption by the γ-CsPbI$_3$ layer. Correspondingly, the predicted J$_{SC}$ from the EQE curves is increased in the PHJ devices and is in a good agreement with the measured J$_{SC}$ (Supplementary Table 2). To confirm that this increase originates from the PHJ structure, we performed optical simulations, the results of which are shown in Supplementary Fig. 11. The excellent agreement in the spectral features of the simulated and experimentally measured EQE spectra confirms that the increased photocurrent originates from the increased light absorption enabled by the addition of the γ-CsPbI$_3$. The high photocurrents also confirm that the small offset in the conduction band position across the PHJ (0.05 eV) does not hinder efficient electron transport from the β-CsPbI$_3$ to the γ-CsPbI$_3$ and its extraction at the PCBM layer. Importantly, the same improvement in the photovoltaic performance cannot be achieved by simply increasing the thickness of the SP photovoltaic devices. We found that the optimal thickness for the β-CsPbI$_3$ SP device is 370 nm (Supplementary Fig. 12), and attempting to make the active layer thicker results in a loss of performance. The maximum thickness that can be achieved (420 nm) is limited by the solubility of the perovskite precursors, thus not allowing to reach the optimum thickness of the PHJ devices (i.e. 470 nm). However, considering that the 420 nm devices show an inferior performance to that of the 370 nm, it is clear that a further increase would not lead to a better performance. This is likely suggested to a poorer microstructure of the thicker β-CsPbI$_3$ devices. Similarly, in our previous work we found the optimal thickness of γ-CsPbI$_3$ SP devices to be 500 nm,[39] which is close to the 470 nm thickness of the PHJ100 devices, yet the SP γ-CsPbI$_3$ devices show a significantly lower photovoltaic performance, due to its larger bandgap and consequently lower J$_{SC}$, highlighting that the PHJ structure is beneficial over each of the SP devices on their own.

The observation that the PHJ200 devices show a slightly reduced performance in comparison to the PHJ100 is likely associated with the poorer microstructure of the 200 nm thick γ-CsPbI$_3$ layers (Fig. 2b). As described above, the surface morphology becomes more disordered with a myriad of grains with a range of shapes and sizes, suggesting that the templating effect by the underlying β-CsPbI$_3$ becomes weaker as the γ-CsPbI$_3$ becomes very thick. Such a disordered microstructure may lead to additional defects, evidenced by a reduction in the PL lifetime (Fig. 3d) and a reduced V$_{OC}$ of the PHJ200 devices.

Current-density-voltage (J-V) characteristics of representative devices of each kind are displayed in Fig. 4f and their photovoltaic parameters are summarised in Table 1. It is noticeable that the β-CsPbI$_3$ devices exhibit non-negligible hysteresis, which is significantly suppressed in both the γ-CsPbI$_3$ SP and PHJ devices. To quantify this, we defined the hysteresis index, (PCE$_{rev}$-

PCE$_{for}$)/PCE$_{rev}$, which is displayed in Fig. 4g for the devices whose performance is shown in Fig. 4a-d. Consistently, the hysteresis index is lower in the PHJ devices than in the β-CsPbI$_3$ solar cells.

Fig. 4h, we summarise the PCE evolution of CsPbI$_3$-based solar cells since their first introduction in 2014.[28,29,31,33,36,37,43–49] The performance of the inverted architecture based PHJ100 is comparable with the standard architecture CsPbI$_3$ solar cells, thus demonstrating the enormous potential of the phase heterojunction solar cell concept.

To further investigate the origin of the improved photovoltaic performance, we focused on the characterisation of the SP β-CsPbI$_3$ solar cell and the PHJ100 device. Light intensity dependent V$_{OC}$ measurements (Fig. 5a) show that the PHJ100 solar cell exhibit an overall lower trap-assisted recombination, as the slope measured for the reference β-CsPbI$_3$ device (1.703 kT q$^{-1}$) is reduced to 1.163 kT q$^{-1}$ for the PHJ devices. While it is not trivial to ascribe charge recombination mechanisms from light intensity dependent V$_{OC}$ and ideality factor measurements in perovskite solar cells[50], these measurements are supported by direct characterisation of the sub-bandgap trap density in the β-CsPbI$_3$ and PHJ100 films by photothermal deflection spectroscopy (PDS). The PDS spectra (Fig. 5b) show that the PHJ100 structure results in a significantly reduced energetic disorder, corresponding to an Urbach energy of 19.8 eV, in comparison to 33.27 eV for the SP β-CsPbI$_3$ thin films. The reduction in the trap density of the PHJ100 structure results in a suppression of non-radiative recombination of losses as can be directly visualised by measuring the solar cells as light-emitting diodes. The electroluminescence quantum yield (ELQY) of the PHJ100 devices is approximately one order of magnitude higher than the SP β-CsPbI$_3$ solar cells (Fig. 5c).

To explore the origin of the reduced hysteresis in the PHJ devices, we characterised the temperature dependent conductivity of the SP β-CsPbI$_3$ and PHJ100 structures integrated in lateral devices, shown in the inset of Fig. 5d. The measurements allow to extract the value of the activation energy ($E_a$) for ion migration by using the Nernst-Einstein equation $\sigma(T) = \frac{\sigma_0}{T} e^{-\frac{E_a}{kT}}$ and have been previously utilized to characterise ion migration in perovskite solar cells[51,52]. We find that the $E_a$ of the SP β-CsPbI$_3$ (0.79 eV) is significantly increased in the case of PHJ100 ($E_a$=1.66 eV). Moreover, the threshold temperature at which ion conduction starts to dominate the total current is increased from 260 K to 273 K. Both of these factors point towards a suppressed ion migration in the PHJ structures, consistent with the significantly reduced hysteresis observed in Fig. 4g.

To evaluate the stability of the PHJ devices, we monitored their performance upon storage under dim light conditions (Fig. 5g). The PHJ100 devices showed an improvement over the first 50 days, reaching a maximum performance of 20.1% (Supplementary Fig. 13). A gradual increase in the photovoltaic performance with storage time has been reported in literature with several different mechanisms proposed[53,54]. While the elucidation of the exact mechanisms at play in our case is beyond the scope of this study, we consistently observed that stored PHJ devices outperform those measured directly after fabrication (Supplementary Fig. 14). The performance remains stable up to 100 days and is only slightly reduced after 180 days. Importantly, throughout the stability study, the PHJ devices consistently exhibited a superior performance to the β-CsPbI$_3$ solar cells. Moreover, we monitored the evolution of the PCE of the devices upon continuous illumination under 1 sun conditions (Fig. 5h). Both types of devices maintained their performance for approximately 400 h, followed by a significant decrease in performance. The fact that both devices exhibit a similar lifetime suggests that their degradation is linked to the stability of the β-CsPbI$_3$, and the fact that over the first 400 h the performance of the PHJ device remained superior to the SP counterpart suggests that the PHJ structure retained its stability for this duration. To further confirm that the PHJ structure is maintained upon illumination, we fabricated PHJ layers with different thicknesses of γ-CsPbI$_3$ and exposed them to 1 sun illumination for 100h. Probing the optical absorption of the PHJ films reveals that it remained unchanged as compared to that of the freshly prepared PHJs, in particular the PHJ100 and PHJ200 structures clearly show the contributions of both the β-CsPbI$_3$ and γ-CsPbI$_3$ phases in their absorption spectra (Supplementary Fig. 15). Furthermore, examining the XRD patters of PHJ100 samples exposed to either 1sun illumination for 100h or thermally annealed at 85 ℃ 15h, shows clearly contributions from both the β-CsPbI$_3$ and γ-CsPbI$_3$ phases, just as is observed for the freshly prepared samples (Supplementary Fig. 16). While there is no doubt that further improvements to device stability are necessary, our initial results are promising since they suggest that the PHJ structure itself is maintained upon prolonged illumination. Future development of degradation mitigation strategies will certainly enhance the device stability further; however, they go beyond the scope of the current study in which the PHJ concept is demonstrated.

**Conclusions**

To summarise, in this work we propose a concept for photovoltaics exploiting the phenomenon of polymorphism of crystalline materials. We demonstrate this concept by fabricating γ-CsPbI$_3$/*β*-CsPbI$_3$ PHJ solar cells, which show superior performance to either of the SP devices. Our findings offer a promising approach towards highly efficient photovoltaic devices that might in the future be realised also via other classes of polymorphic materials, for example, organic materials or other perovskite compositions.

**Methods**

Indium tin oxide (ITO) coated glass substrates were bought from PsiOTech Ltd. [2-(3,6-Dimethoxy-9H-carbazol-9-yl)ethyl]phosphonic Acid (Meo-2PACz) was purchased from TCI. Phenyl-C$_{61}$-butyric acid methyl ester (PC$_{61}$BM) (99.5%) and bathocuproine (BCP, 99.99%, trace metals basis) were obtained from Solenne BV and Sigma Aldrich, respectively. TiCl (99.9%) were obtained from Alfa Aesar. Spiro-OMeTAD (99%, HPLC), TBP (98%), and Li-TFSI were obtained from Sigma Aldrich. All solvents were purchased from Sigma-Aldrich. For solution processed β-CsPbI$_3$, PbI$_2$ (99.999%, trace metals basis), CsI (99.999%, trace metals basis) were obtained from TCI. The PbI$_2$·xDMAI was prepared by using PbI$_2$ (1.2 g) dissolved in 2.5 mL Anhydrous dimethylformamide (DMF) at 80 °C under active stirring for 30 min in air atmosphere. Immediately after, 10 mL of hydroiodic acid were added to the solution with stirring for 24 h at 80 °C. The precipitates were filtered then centrifuged and washed with several times copious diethyl ether and ethyl alcohol to remove the residuary solvent, respectively. The collected powders are dried in a vacuum oven at 70 °C for 24 h. For evaporated γ-CsPbI$_3$, PbI$_2$ (99.999%, trace metals basis), CsI (99.999%, trace metals basis) and Phenethylammonium iodide (PEAI, 98%) were obtained from Sigma Aldrich. All materials were used as received without purification.

**Perovskite Film Deposition**

Substrates for film deposition were ultrasonically cleaned with 2 % Hellmanex© detergent, deionized water, acetone, and isopropanol followed by 15 min oxygen plasma treatment. In a drybox (RH < 1 %), Meo-2PACz (3 mg/ml in ethanol) was spin-coated on the clean substrates with 4000 rpm for 30 s and annealed at 100 °C 10 min. For n-i-p devices, laser-patterned FTO/glass substrate (2.9 cm × 2.9 cm) was immersed into aqueous solution of TiCl (4.5 mL TiCl in 200 mL H$_2$O) at 70 °C for 1 h, followed by annealing at 200 °C for 0.5 h to obtain a compact TiO$_2$ electron transport layer (ETL). The TiO$_2$ was then cleaned using an UV-Ozone cleaner for 12 min. The CsPbI$_3$ precursor solutions was prepared by dissolving 0.8025 g PbI$_2$·xDMAI, 0.2107 g PbI$_2$ and 0.4159 g CsI into the solution of 2 mL DMF and dimethyl sulfoxide (DMSO) (v/v, 9:1) under active stirring for 12 h at 60 °C. The CsPbI$_3$ solution was spin-coated at 1000 rpm 10 s and 4500 rpm 30 s. Then the samples were annealed at 180 °C for 15 min. Next the samples were transferred into a vacuum chamber (CreaPhys GmbH, Germany) to deposit γ-CsPbI$_3$ with desired thickness following previously reported procedures.[39] Next, the samples were transferred into a nitrogen-filled glove box (GS), where PC$_{61}$BM (20 mg/ml dissolved in chlorobenzene) was dynamically spin-coated at 2000 rpm for 30 s followed by a 3 min annealing at 100 °C. Finally, a BCP (0.5 mg/ml dissolved in isopropanol) hole-blocking layer was spin-coated at 4000 rpm for 30 s, following by an 80 nm thermally evaporated Ag cathode (Mantis evaporator, base pressure of 10$^{-7}$ mbar). For n-i-p devices, 6.0 μL of Spiro-OMeTAD solution prepared by dissolving 90 mg Spiro-OMeTAD in 1.0 mL chlorobenzene containing 36 μL TBP and 22 μL of Li-TFSI was spin-cast on the perovskite layer at 1500 rpm 40s. Then, the samples were transferred into a drybox for the oxidation of Spiro-OMeTAD for an entire night. Finally, a 100 nm thick gold electrode was deposited by thermal evaporation to complete the cell. The as-prepared photovoltaic devices were sealed/encapsulated in a glovebox using an UV-hardened epoxy glue and a transparent clean encapsulation glass.

**Photovoltaic Device Characterization**

EQE spectra of the devices were recorded using a monochromatised light of a halogen lamp from 400 nm to 800 nm and the reference spectra were calibrated using an NIST-traceable Si diode (Thorlabs). J-V characteristics of solar cells under a solar simulator (Abet Sun 3000 Class AAA solar simulator, AM 1.5 conditions) were recorded at room temperature in the ambient using a computer controlled Keithley 2450 source measure unit. The incident light intensity was calibrated via a Si reference cell (NIST traceable, VLSI) and tuned by measuring the spectral mismatch factor between the real solar spectrum, the spectral response of the reference cell and the perovskite devices. All devices were scanned from short circuit to forward bias (1.2 V) and reverse with a rate of 0.025 V s$^{-1}$. No treatment was applied prior to measurements. The active area for all devices was 4.5 mm$^2$ defined by thermal evaporation of the Ag electrodes.

**Ion Migration for Activation Energy Characterization**

The activation energy of ion migration was extracted from the dependence of the conductivity of the CsPbI$_3$ films on temperature. In short, we used a lateral structure device which consists of two Au (50 nm) electrodes with a length 111 mm and spacing gap of 0.2 mm deposited on surface of CsPbI$_3$ films with and without hetero-phase structure. In measurement, an electric field of 0.3 Vμm$^{-1}$ was applied for all the devices. The measurements were conducted with a home-made set-up. To change the cell temperature, the devices were placed in vacuum on a copper block, which was connected to a Peltier element (Peltron), controlled by a BelektroniG HAT Control device. For each conductivity, the current through the devices were stabilized for 5 minutes when an objective temperature was reached, before the current measurement was performed. A semiconductor analyzer (Keithley SMU2635A) was used for the current measurement. The measurement equipment was controlled with the software SweepMe! (https://sweep-me.net).

**Scanning-Electron Microscopy**

A SEM (Gemini 500, (ZEISS, Oberkochen, Germany)) with an acceleration voltage of 3 kV under 5-6×10$^{-4}$ mbar was utilized to obtain the surface and cross-sectional SEM images using the in-lens mode.

**X-Ray Diffraction**

XRD patterns were measured in ambient air by using a Bruker Advance D8 diffractometer equipped with a 1.6 kW Cu-Anode (λ = 1.54060 Å) and a LYNXEYE_XE_T 1D-Mode detector. The scans (2theta-Omega mode, 2θ = 10°-40°, step size 0.01°, 0.1 s/step) were measured in a parallel beam geometry with a height limiting slit of 0.2 mm. For grazing incidence XRD, the parameters of scans are 2theta mode, 2θ = 13.5°-15°, 28°-29.6° step size 0.01°, 0.5 s/step, The incidence angle (Ω) was fixed at 0.4°, 0.5°, 1°, 2°, 5°, respectively.

**UV-vis Absorption and Photoluminescence Measurements**

A Shimadzu UV-3100 spectrometer was utilized in the ultraviolet−visible (UV−vis) absorbance spectra recording. PL measurements were performed using a CW laser (447 nm, 10 mW, Coherent) as the excitation source. The PL signal was collected using a NIR spectrometer (OceanOptics). All samples were with encapsulated to prevent the decomposition and enable the all measurements to be were carried out in ambient air at room temperature.

**Photothermal Deflection Spectroscopy (PDS)**

PDS measurements were performed following the procedure described in our previous work.[55–57] The quartz substrate with perovskite film were mounted in the signal enhancing liquid (Fluorinert FC-770) filled quartz cuvette inside a N2 filled glovebox. Then, the samples were excited using a tunable, chopped, monochromatic light source (150W xenon short arc lamp with a Cornerstone monochromator) and probed using a laser beam (635nm diode laser, Thorlabs) propagating parallel to the surface of the sample. The heat generated through the absorption of light changes the refractive index of the Fluorinert liquid, resulting in the deflection of the laser beam. This deflection was measured using a position sensitive-detector (Thorlabs, PDP90A) and a lock-in amplifier (Amatec SR7230) and is directly correlated to the absorption of the film.

**Time-Correlated Single Photon Counting (TCSPC)**

A TCSPC setup containing of a 375 nm laser diode head (Pico Quant LDHDC375), a PMA Hybrid Detector (PMA Hybrid 40), a TimeHarp platine (all PicoQuant), and a Monochromator SpectraPro HRS-300 (Princeton Instruments) was utilized. Perovskite films on quartz were excited with the 375 nm laser diode. The emission was collected by the PMA hybrid detector. The excitation fluence is around 7.5 nJ/cm$^2$. The lifetimes were evaluated using reconvolution algorithms of FluoFit (PicoQuant).

**Electroluminescence (EL) Measurements**

The devices were mounted in an integrating sphere (Labsphere Inc.). The current–voltage characteristics were recorded using a Keithley 2450 source-measure unit. At the same time, a scientific grade spectrometer (Ocean Optics QE65PRO) was used to record the emission spectra, which were used to calculate the luminance. The optical system (spectrometer, integrating sphere, and coupling optical fiber) was calibrated via a calibrated light source (Ocean Optics HL-2000-CAL).

**Ultra-Violet Photoemission Spectroscopy (UPS) Depth Profiling (DP)**

The samples were transferred to an ultrahigh vacuum chamber (ESCALAB 250Xi), with a base pressure of $2*10^{-10}$ mbar, for UPS DP measurements. UPS measurements were performed using a double-differentially pumped He gas discharge lamp emitting He I radiation (hv=21.22 eV) with a pass energy of 2 eV and a bias of -5 V in order to ensure secondary electron onset detection. The UPS spectra are shown as a function of the binding energy with respect to the Fermi energy. The energy edge of the valence band features is used to determine the valence band level position with respect to the Fermi level. The conduction bands were estimated using the optical gaps of the γ-CsPbI$_3$ and β-CsPbI$_3$. Cluster etching was performed using large Ar clusters generated by the MAGCIS Dual Beam Ion Source (Thermo Scientific) with an energy of 4000 eV.

**Optical Simulation**

Optical simulations are performed using an in-house developed optical simulation tool based on the transfer-matrix-method (TMM). The optical constants of perovskite layers are obtained by variable angle spectroscopic ellipsometry utilizing an EP4 imaging ellipsometer (Accurion GmbH,

Germany). An effective isotropic optical model is applied in order to extract the complex refractive indices. The dispersion of both perovskite layer types is modelled by a superposition of four Tauc-Lorentz oscillators resulting in unbiased root-mean-square errors of < 2.

**Data availability**

All data generated or analysed during this study are included in the published article and its Supplementary Information and Source Data files.

**Acknowledgements**

This project has received funding from the European Research Council (ERC) under the European Union's Horizon 2020 research and innovation programme (ERC Grant Agreement n° 714067, ENERGYMAPS) and the Deutsche Forschungsgemeinschaft (DFG) in the framework of the Special Priority Program (SPP 2196) project PERFECT PVs (#424216076). R.J. and Z.Z. are grateful for the financial support by the China Scholarship Council (Scholarship #201806070145 and #201806750012, respectively). R. J. thanks Prof. Yuan Liu, Dr. Yungui Li, Dr. Yifan Zheng, Dr. Xiangkun Jia, Shen Xing for useful discussions and Marielle Deconinck for help with the UPS depth profiling measurements. Y. V. thanks Prof. Nir Tessler and Dr. Artem A. Bakulin for fruitful discussions. We also acknowledge support from Prof. Sebastian Reineke in financing C. H..

**Author contributions**

R.J. conceived the idea, designed experiments and analysed data. R.J. and Z. Z. fabricated perovskite thin films and photovoltaic devices and characterized the SEM for the perovskite thin films. R.J. synthesized $PbI_2 \cdot xDMAI$ precursor, carried out the XRD, EL, conductivity measurements, optical modelling and photovoltaic devices testing. C. H. measured n, k values for optical modelling. R. B. performed the PDS measurements and analysis. Y. H. performed the UPS DP measurements and analysis. R. J. and F. P. performed the PL measurements and analysis. Y.V. supervised and guided the work. R. J. and Y. V. wrote the manuscript which has been commented on and edited by all co-authors.

**Competing interests**

The authors declare no competing interests.

**Tables:**

**Table 1**: Photovoltaic performance parameters of devices whose J-V curves are displayed in Fig. 4f.

|  | $V_{OC}$ (V) | $J_{SC}$ (mA/cm$^2$) | FF (%) | PCE (%) | $V_{OC}$ (V) | $J_{SC}$ | FF (%) | PCE (%) | Hysteresis Index (%) |
|---|---|---|---|---|---|---|---|---|---|
|  | Forward | | | | Reverse | | | | |
| β-CsPbI$_3$ | 0.99 | 18.53 | 70.24 | 12.91 | 1.02 | 18.53 | 76.24 | 14.38 | 0.10 |
| PHJ2 | 1.03 | 18.34 | 71.53 | 13.54 | 1.06 | 18.34 | 79.41 | 15.41 | 0.12 |
| PHJ5 | 1.04 | 18.18 | 72.86 | 13.83 | 1.08 | 18.18 | 79.62 | 15.66 | 0.12 |
| PHJ10 | 1.09 | 18.94 | 79.51 | 16.46 | 1.10 | 18.94 | 82.40 | 17.13 | 0.04 |
| PHJ20 | 1.10 | 19.44 | 78.63 | 16.77 | 1.10 | 19.44 | 83.50 | 17.85 | 0.06 |
| PHJ50 | 1.12 | 19.71 | 81.13 | 17.89 | 1.13 | 19.71 | 82.71 | 18.35 | 0.03 |
| PHJ100 | 1.15 | 20.07 | 82.07 | 18.94 | 1.15 | 20.07 | 82.76 | 19.10 | 0.01 |
| PHJ200 | 1.10 | 19.16 | 77.79 | 16.41 | 1.10 | 19.16 | 84.30 | 17.80 | 0.09 |
| γ-CsPbI$_3$ | 1.12 | 17.58 | 75.36 | 14.78 | 1.12 | 17.58 | 76.54 | 15.05 | 0.02 |

**Figures:**

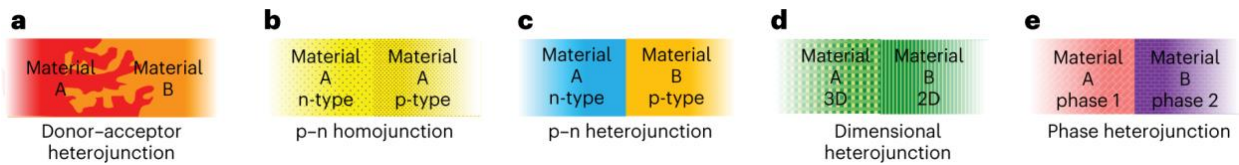

a. Donor–acceptor heterojunction
b. p–n homojunction
c. p–n heterojunction
d. Dimensional heterojunction
e. Phase heterojunction

**Figure 1**: **Schematic illustration of different types of heterojunctions in photovoltaics and PHJ fabrication procedure.** (a) Donor-acceptor heterojunction, (b) PN homojunction, (c) PN heterojunction, (d) 3D/2D dimensional heterojunction and (e) phase heterojunction.

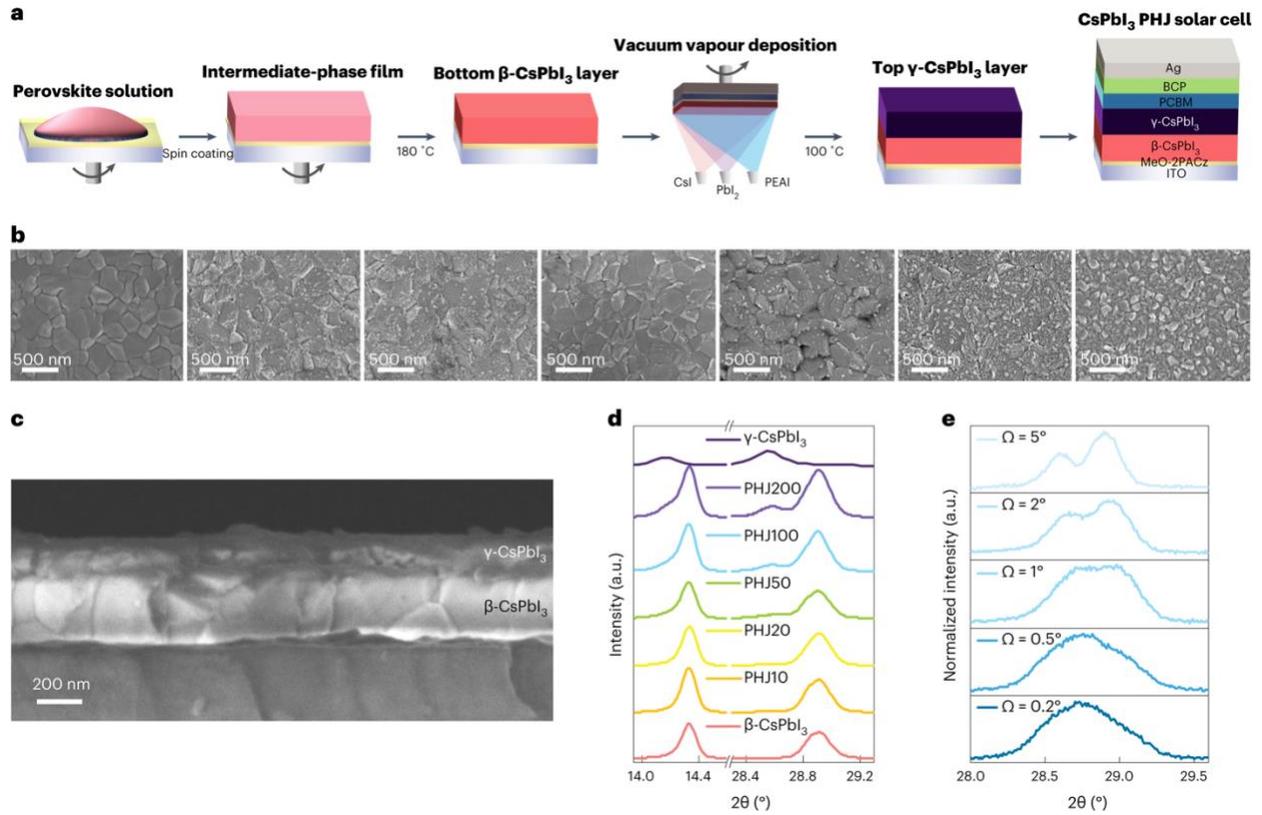

**Figure 2**. **Structural and microstructural characterisation of PHJs.** (a) Schematic illustration of the fabrication procedure of the β-CsPbI$_3$ / γ-CsPbI$_3$ PHJ. (b) Surface SEM images of β-CsPbI$_3$, γ-CsPbI$_3$ and PHJ layers with varying thickness. (c) Cross-sectional SEM of PHJ100. (d) XRD of β-CsPbI$_3$, γ-CsPbI$_3$ and PHJ layers with varying thickness. (e) XRD of PHJ100 measured at different incidence angles.

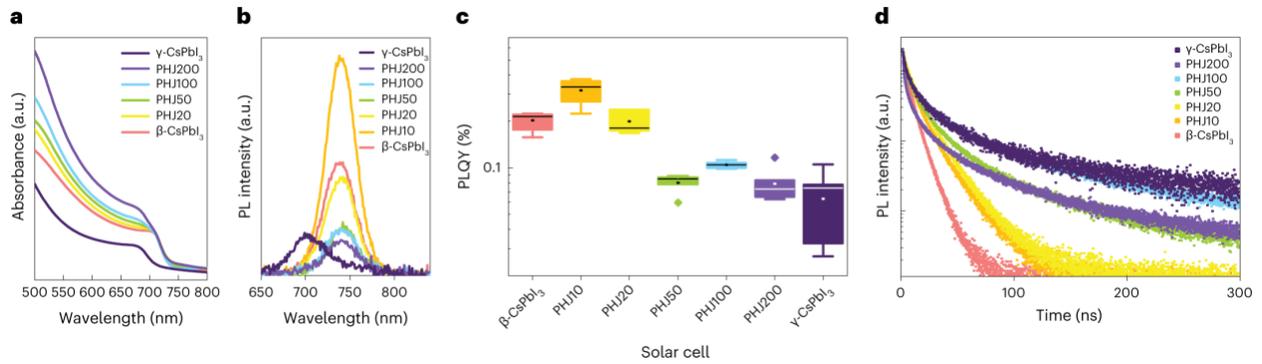

**Figure 3: Optical properties of SP and PHJ films.** (a) UV, (b) PL, (c) PLQY and (d) PL lifetime measured on SP γ-CsPbI$_3$ and β-CsPbI$_3$ and PHJ with varying layer thickness. The box plot in panel c displays the mean, median line, 25%~75% box limits with 1.5x interquartile range whiskers. On average 6 samples of each type were measured to obtain the PLQY data displayed in panel c.

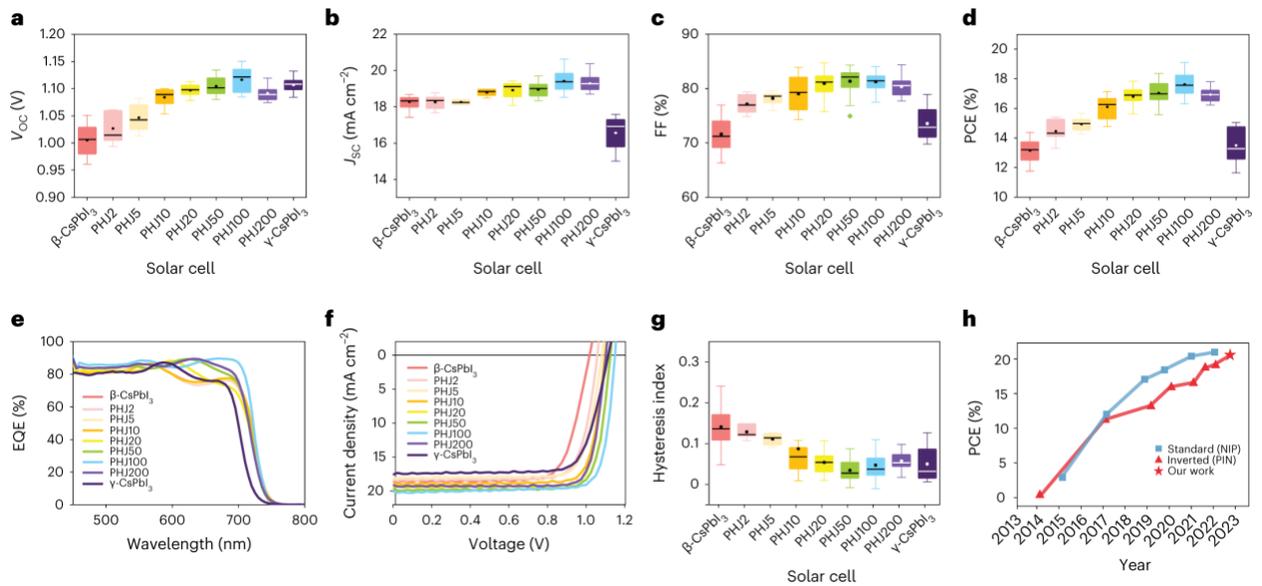

**Figure 4**. **Photovoltaic performance of SP and PHJ solar cells.** Distribution of (a) V$_{OC}$, (b) J$_{SC}$, (c) FF, (d) PCE parameters (e) EQE spectra (f) J-V characteristics (g) hysteresis index of SP and PHJ solar cells with varying thickness. (h) comparison of PCEs of standard and inverted architecture CsPbI$_3$ solar cells (data taken from references [28,29,31,33,36,37,43–49]). The curves drawn on top of data are presented as guides to the eye. The box plots in panels a-d and g display the mean, median line, 25%~75% box limits with 1.5x interquartile range whiskers. The number of samples of each column of panels a, b, c, d and g is 64, 7, 8, 23, 30, 46, 120, 39, 74, respectively. 21 samples from the same batch for 'β-CsPbI$_3$' in a, b, c, d, 21 data points of the same batch behind the average values are already provided in the Supplementary Figure 12 as '370nm'. From 'β-CsPbI$_3$', 'γ-CsPbI$_3$', 'PHJ100' in d, 18 data points of the same batch behind the average values are already provided in the Supplementary Figure 14.

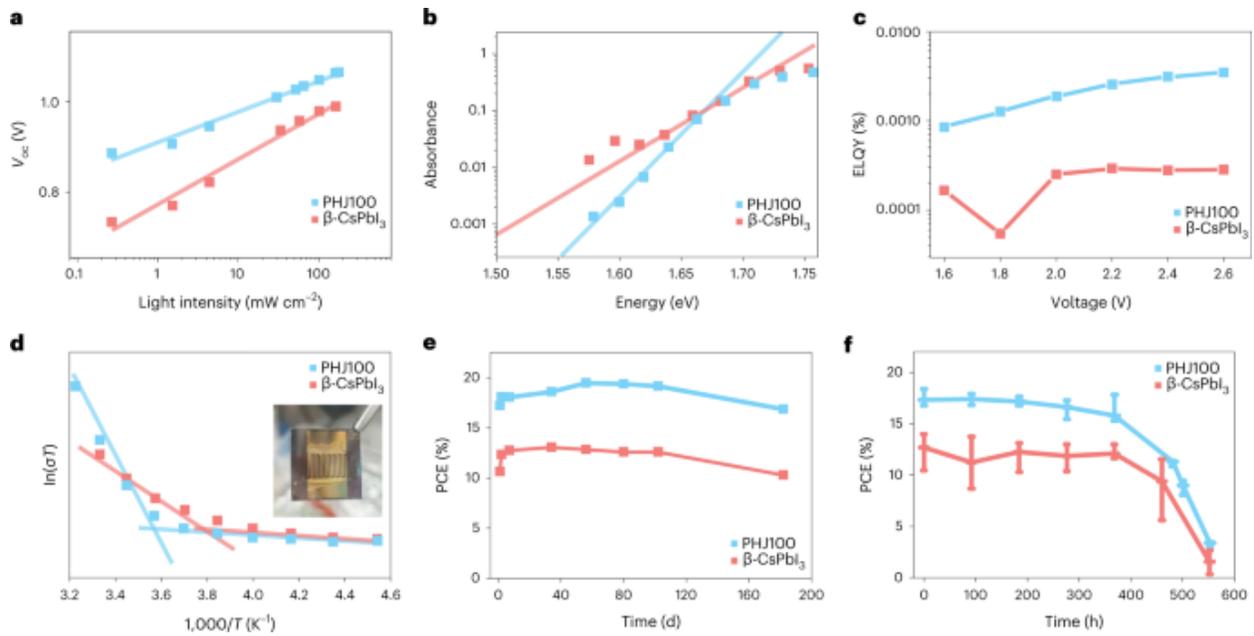

**Figure 5. Origin of photovoltaic performance improvement and device stability.** (a) Light intensity dependent V$_{OC}$ showing ideality factors (n) of 1.703 and 1.163 for SP β-CsPbI$_3$ and PHJ100, respectively. (b) PDS spectra showing the Urbach energies ($E_U$) of 33.27 meV and 19.80 meV for SP β-CsPbI$_3$ and PHJ100, respectively. (c) ELQY (d) Temperature dependent conductivity showing ion migration activation energies (Ea) of 0.79 eV and 1.66 eV, and the transition temperatures of 260 K and 273 K for SP β-CsPbI$_3$ and PHJ100, respectively. The inset in d shows the conductivity device with lateral structure consisting of two Au (50 nm) electrodes

with a length 111 mm and spacing gap of 0.2 mm deposited on the surface of perovskite films. (e) Storage stability of PCE. Devices stored in nitrogen atmosphere at 25°C and under dark conditions. (f) Continuous illunination stability of encapsulated SP β-CsPbI$_3$ and PHJ100 solar cells. Up to 550 h stability was tested under 40% RH, 25°C, constant xenon-lamp-simulated solar illumination (100 mW cm$^{-2}$) without a UV filter. The fitting lines in a, b, and d are obtained by linear fitting of data points. The panel f displays the mean, with 1.5x interquartile range whiskers.

**References**

1. Bach, U. *et al.* Solid-state dye-sensitized mesoporous TiO$_2$ solar cells with high photon-to-electron conversion efficiencies. *Nature* **395**, 583–585 (1998).

2. Peumans, P., Uchida, S. & Forrest, S. R. Efficient bulk heterojunction photovoltaic cells using small-molecular-weight organic thin films. *Nature* **425**, 158–162 (2003).

3. Pattantyus-Abraham, A. G. *et al.* Depleted-Heterojunction Colloidal Quantum Dot Solar Cells. *ACS Nano* **4**, 3374–3380 (2010).

4. Heo, J. H. *et al.* Efficient inorganic–organic hybrid heterojunction solar cells containing perovskite compound and polymeric hole conductors. *Nature Photon* **7**, 486–491 (2013).

5. Liu, M., Johnston, M. B. & Snaith, H. J. Efficient planar heterojunction perovskite solar cells by vapour deposition. *Nature* **501**, 395–398 (2013).

6. Yu, G., Gao, J., Hummelen, J. C., Wudl, F. & Heeger, A. J. Polymer Photovoltaic Cells: Enhanced Efficiencies via a Network of Internal Donor-Acceptor Heterojunctions. *Science* **270**, 1789-1791 (1995).

7. Wadsworth, A., Hamid, Z., Kosco, J., Gasparini, N. & McCulloch, I. The Bulk Heterojunction in Organic Photovoltaic, Photodetector, and Photocatalytic Applications. *Advanced Materials* **32**, 2001763 (2020).

8. Yuan, J. *et al.* Effects of energetic disorder in bulk heterojunction organic solar cells. *Energy Environ. Sci.* **15**, 2806–2818 (2022).

9. Chen, J. *et al.* Hole (donor) and electron (acceptor) transporting organic semiconductors for bulk-heterojunction solar cells. *EnergyChem* **2**, 100042 (2020).

10. Shah, A., Torres, P., Tscharner, R., Wyrsch, N. & Keppner, H. Photovoltaic Technology: The Case for Thin-Film Solar Cells. *Science* **285**, 692-698 (1999).

11. Möller, H. J. Semiconductors for solar cells. *Semiconductors for solar cells* 343–343 (1993).

12. Heske, C. *et al.* Observation of intermixing at the buried CdS/Cu(In,Ga)Se2 thin film solar cell heterojunction. *Appl. Phys. Lett.* **74**, 1451–1453 (1999).

13. Fedotov, Y. A., Zased, V. S. & Matson, É. A. Electrical Properties of α (Ge) — GaAs Heterojunctions. in *Physics of p-n Junctions and Semiconductor Devices* (eds. Ryvkin, S. M. & Shmartsev, Y. V.) 104–106 (Springer US, 1971). doi:10.1007/978-1-4757-1232-2_22.

14. McGott, D. L. *et al.* 3D/2D passivation as a secret to success for polycrystalline thin-film solar cells. *Joule* **5**, 1057–1073 (2021).

15. Bernstein, J. *Polymorphism in Molecular Crystals 2e*. (Oxford University Press, 2020).

16. Gentili, D., Gazzano, M., Melucci, M., Jones, D. & Cavallini, M. Polymorphism as an additional functionality of materials for technological applications at surfaces and interfaces. *Chemical Society Reviews* **48**, 2502–2517 (2019).

17. Caira, M. R. Crystalline polymorphism of organic compounds. in *Design of Organic Solids* (ed. Weber, E.) vol. 198 163–208 (Springer-Verlag Berlin, 1998).

18. Bu, R., Li, H. & Zhang, C. Polymorphic Transition in Traditional Energetic Materials: Influencing Factors and Effects on Structure, Property, and Performance. *Cryst. Growth Des.* **20**, 3561–3576 (2020).

19. Sood, S. & Gouma, P. Polymorphism in nanocrystalline binary metal oxides. *Nanomater. Energy* **2**, 82–96 (2013).

20. Nogueira, B. A., Castiglioni, C. & Fausto, R. Color polymorphism in organic crystals. *Commun Chem* **3**, 34 (2020).

21. Schmidt-Mende, L. *et al.* Roadmap on organic–inorganic hybrid perovskite semiconductors and devices. *APL Materials* **9**, 109202 (2021).

22. Min, H. *et al.* Perovskite solar cells with atomically coherent interlayers on $SnO_2$ electrodes. *Nature* **598**, 444–450 (2021).

23. Chen, S. *et al.* Stabilizing perovskite-substrate interfaces for high-performance perovskite modules. *Science* **373**, 902–907 (2021).

24. Degani M. *et al.* 23.7% Efficient inverted perovskite solar cells by dual interfacial modification. *Science Advances* **7**, eabj7930 (2021).

25. Bube, R. H. *et al.* Photovoltaic energy conversion with n-CdS—p-CdTe heterojunctions and other II–VI junctions. IEEE Trans. Electron Devices 24, 487–492 (1977).

26. Sutton, R. J. *et al.* Cubic or Orthorhombic? Revealing the Crystal Structure of Metastable Black-Phase $CsPbI_3$ by Theory and Experiment. *ACS Energy Lett.* **3**, 1787–1794 (2018).

27. Marronnier, A. *et al.* Anharmonicity and Disorder in the Black Phases of Cesium Lead Iodide Used for Stable Inorganic Perovskite Solar Cells. *ACS Nano* **12**, 3477–3486 (2018).

28. E. Eperon, G. *et al.* Inorganic caesium lead iodide perovskite solar cells. *Journal of Materials Chemistry A* **3**, 19688–19695 (2015).

29. Wang, Y. *et al.* Thermodynamically stabilized beta-CsPbI$_3$-based perovskite solar cells with efficiencies > 18%. *Science* **365**, 591-595 (2019).

30. Stoumpos, C. C. & Kanatzidis, M. G. The Renaissance of Halide Perovskites and Their Evolution as Emerging Semiconductors. *Accounts of Chemical Research* **48**, 2791–2802 (2015).

31. Wang, Y., Zhang, T., Kan, M. & Zhao, Y. Bifunctional Stabilization of All-Inorganic α-CsPbI$_3$ Perovskite for 17% Efficiency Photovoltaics. *J. Am. Chem. Soc.* **140**, 12345–12348 (2018).

32. Ye, Q. *et al.* Stabilizing γ-CsPbI$_3$ Perovskite via Phenylethylammonium for Efficient Solar Cells with Open-Circuit Voltage over 1.3 V. *Small* **16**, 2005246 (2020).

33. Wang, Y. *et al.* The Role of Dimethylammonium Iodide in CsPbI$_3$ Perovskite Fabrication: Additive or Dopant? *Angewandte Chemie International Edition* **58**, 16691–16696 (2019).

34. Chang, X. *et al.* Printable CsPbI$_3$ Perovskite Solar Cells with PCE of 19% via an Additive Strategy. *Advanced Materials* **32**, 2001243 (2020).

35. Du, Y. *et al.* Ionic Liquid Treatment for Highest-Efficiency Ambient Printed Stable All-Inorganic CsPbI$_3$ Perovskite Solar Cells. *Advanced Materials* **n/a**, 2106750.

36. Yoon, S. M. *et al.* Surface Engineering of Ambient-Air-Processed Cesium Lead Triiodide Layers for Efficient Solar Cells. *Joule* **5**, 183–196 (2021).

37. Fu, S. *et al.* Humidity-Assisted Chlorination with Solid Protection Strategy for Efficient Air-Fabricated Inverted CsPbI$_3$ Perovskite Solar Cells. *ACS Energy Lett.* **6**, 3661–3668 (2021).

38. Wang, Y., Chen, Y., Zhang, T., Wang, X. & Zhao, Y. Chemically Stable Black Phase CsPbI$_3$ Inorganic Perovskites for High-Efficiency Photovoltaics. *Advanced Materials* **32**, 2001025 (2020).

39. Zhang, Z. *et al.* Efficient Thermally Evaporated γ-CsPbI$_3$ Perovskite Solar Cells. *Advanced Energy Materials* **11**, 2100299 (2021).

40. Wang, K. *et al.* In-Situ Hot Oxygen Cleansing and Passivation for All-Inorganic Perovskite Solar Cells Deposited in Ambient to Breakthrough 19% Efficiency. *Advanced Functional Materials* **31**, 2101568 (2021).

41. Zhao, B. *et al.* Thermodynamically Stable Orthorhombic γ-CsPbI$_3$ Thin Films for High-Performance Photovoltaics. *J. Am. Chem. Soc.* **140**, 11716–11725 (2018).

42. Tessler, N. Adding 0.2 V to the open circuit voltage of organic solar cells by enhancing the built-in potential. *Journal of Applied Physics* **118**, 215501 (2015).

43. Choi, H. *et al.* Cesium-doped methylammonium lead iodide perovskite light absorber for hybrid solar cells. *Nano Energy* **7**, 80–85 (2014).

44. Wu, T. *et al.* Efficient and Stable CsPbI$_3$ Solar Cells via Regulating Lattice Distortion with Surface Organic Terminal Groups. *Advanced Materials* **31**, 1900605 (2019).

45. Wang, Q. *et al.* Stabilizing the α-Phase of CsPbI$_3$ Perovskite by Sulfobetaine Zwitterions in One-Step Spin-Coating Films. *Joule* **1**, 371–382 (2017).

46. Fu, S. *et al.* Tailoring In Situ Healing and Stabilizing Post-Treatment Agent for High-Performance Inverted CsPbI$_3$ Perovskite Solar Cells with Efficiency of 16.67%. *ACS Energy Lett.* **5**, 3314–3321 (2020).

47. Wang, J. *et al.* Highly efficient all-inorganic perovskite solar cells with suppressed non-radiative recombination by a Lewis base. *Nat Commun* **11**, 177 (2020).

48. Zhang, T. *et al.* Bication lead iodide 2D perovskite component to stabilize inorganic a-CsPbI$_3$ perovskite phase for high-efficiency solar cells. *SCIENCE ADVANCES* **3**, e1700841 (2017).

49. Tan, S. *et al.* Temperature-Reliable Low-Dimensional Perovskites Passivated Black-Phase CsPbI3 toward Stable and Efficient Photovoltaics. *Angewandte Chemie* **134**, e202201300 (2022).

50. Le Corre, V. M., Sherkar, T. S., Koopmans, M. & Koster, L. J. A. Identification of the dominant recombination process for perovskite solar cells based on machine learning. *Cell Reports Physical Science* **2**, 100346 (2021).

51. Zhao, J. *et al.* Strained hybrid perovskite thin films and their impact on the intrinsic stability of perovskite solar cells. *Science Advances* **3**, eaao5616 (2017).

52. Yang, S. *et al.* Stabilizing halide perovskite surfaces for solar cell operation with wide-bandgap lead oxysalts. *Science* **365**, 473 (2019).

53. Cho, Y. *et al.* Immediate and Temporal Enhancement of Power Conversion Efficiency in Surface-Passivated Perovskite Solar Cells. *ACS Appl. Mater. Interfaces* **13**, 39178–39185 (2021).

54. Moghadamzadeh, S. *et al.* Spontaneous enhancement of the stable power conversion efficiency in perovskite solar cells. *J. Mater. Chem. A* **8**, 670–682 (2020).

55. Fassl, P. *et al.* Fractional deviations in precursor stoichiometry dictate the properties, performance and stability of perovskite photovoltaic devices. *Energy & Environmental Science* **11**, 3380–3391 (2018).

56. Becker-Koch, D. *et al.* Probing charge transfer states at organic and hybrid internal interfaces by photothermal deflection spectroscopy. *J. Phys.: Condens. Matter* **31**, 124001 (2019).

57. Sun, Q. *et al.* Role of Microstructure in Oxygen Induced Photodegradation of Methylammonium Lead Triiodide Perovskite Films. *Advanced Energy Materials* **7**, 1700977 (2017).